\documentclass[twocolumn,showpacs,preprintnumbers,amsmath,amssymb,aps,pra,nobibnotes]{revtex4}
\expandafter\ifx\csname package@font\endcsname\relax\else
\expandafter\expandafter \expandafter\usepackage
\expandafter\expandafter
\expandafter{\csname package@font\endcsname}%
\fi
\DeclareRobustCommand\substyle{\name@idx{document substyle}}%
\DeclareRobustCommand\classoption{\name@idx{document class option}}%
\DeclareRobustCommand\classname{\name@idx{document class}}%
\def\name@idx#1#2{%
 {\ttfamily#2}%
 \index{#2\space#1=\string\ttt{#2}\space#1}\index{#1>#2=\string\ttt{#2}}%
}%

\usepackage{amssymb}
\usepackage{graphicx}
\usepackage{dcolumn}
\usepackage{bm}
\usepackage{amsmath}
\usepackage{amscd}
\usepackage{balance}
\usepackage[all]{xy}

\begin{document}
\title{Exact solution of one class of Maryland model}
\author{Tao Ma}
\affiliation{Department of Modern Physics, University of Science and
Technology of China, Hefei, PRC}
\date{\today}

\begin{abstract}
The Hamiltonian $H$ of one-body Maryland model is defined as the sum
of a linear unperturbed Hamiltonian $H_0$ ($H_{0nn}=n\omega$) and
the interaction $V$, which is a Toeplitz matrix. Maryland model with
a doubly infinite Hilbert space are exactly solved. Special cases of
one-body Maryland model include the original Maryland model (Phys.
Rev. Lett. 49, 833 (1982) and Physica 10D, 369 (1984)), which
describes a quantum kickied linear rotator and single band Bloch
oscillations. Maryland model and single band Bloch oscillations are
the same Hamiltonian in two different representations. A special
case of many-body Maryland model is Luttinger model.

\end{abstract}
 \maketitle

\section{\label{sec:level1}Exact solution}
The Hamiltonian of one-body Maryland model is defined as
\begin{equation}
H=H_0+V.
\end{equation}
Both the unperturbed Hamiltonian $H_0$ and the interaction $V$ are
doubly infinite matrices; the indices $n$ and $m$ of $H_{0nm}$ and
$V_{nm}$ run from $-\infty$ to $\infty$. $H_0$ is a diagonal matrix;
its diagonal matrix elements are
$\{-\infty,\cdots,-2,-1,0,1,2,\cdots,\infty \}\times \omega(t)$. $V$
is a Toeplitz matrix: $V_{nm}=V_{n-m}$.

We want to solve Schrodinger equation
\begin{equation}
i\hbar \frac{\partial}{\partial t} \psi(t)=H \psi(t).
\end{equation}
The unitary operator $U(t,0)$ dictates the dynamics of the system.
\begin{equation}
\psi(t)=U(t,0)\psi(0).
\end{equation}
In the interaction picture,
\begin{equation}
i \hbar \frac{\partial}{\partial t} U_I(t,0)=H_I(t) U_I(t,0).
\end{equation}
where
\begin{equation}
H_I(t)=e^{\frac{i}{\hbar} \int_0^t H_0(t') \, dt'}V
e^{-\frac{i}{\hbar} \int_0^t H_0(t') \, dt'} \\.
\end{equation}
It can be verified, the matrix element of $H_I(t)$ is
\begin{equation}
H_I(t)_{nm}=V_{n-m}e^{\frac{i}{\hbar}(n-m) \int_0^t \omega(t')\,
dt'}.
\end{equation}
$H_I(t)$ is a Laurent matrix. See the Appendix A for a definition of
Laurent matrix. From the theorem of Laurent matrix in the Appendix
A, $[H_I(t), H_I(t')]=0$.

From the scattering theory,
\begin{equation}
U_I(t,0)=P \exp\bigg[-\frac{i}{\hbar} \int_0^t H(t') \, dt'\bigg],
\end{equation}
where $P$ is the time ordering operator. Since $[H_I(t),
H_I(t')]=0$, we can remove the time ordering operator $P$ here.
\begin{equation}
U_I(t,0)=\exp\bigg[-\frac{i}{\hbar} \int_0^t H_I(t') \, dt'\bigg].
\end{equation}
The solution of the Maryland model Eq. $(1)$ is expressed as the
matrix exponential of a Laurent matrix.

In the paper, we use Eq. $(8)$ to solve two special cases of
Maryland model, the original Maryland model \cite{Grempel1982,
Prange1982, Berry1984} and single band Bloch oscillations
\cite{Dunlap1986}. As the first Hamiltonian of the form in Eq. $(1)$
was discussed in \cite{Grempel1982, Prange1982, Berry1984}, the
model Hamiltonian in Eq. $(1)$ is referred as Maryland model.
Another term can be linear Toeplitz system, which emphasizes the
matrix structure of the Hamiltonian and the role of the time
ordering operator.

\section{\label{sec:level1}Connection with Bloch oscillations}
The physical explanation of the Hamiltonian in Eq. $(1)$ can be the
original Maryland model \cite{Grempel1982, Prange1982, Berry1984}.
In the rotator representation \cite{Grempel1982, Prange1982,
Berry1984}, $H_0=-i\hbar \frac{\partial}{\partial \theta} \times
\omega(t)$, $V=V(\theta, t)$ and $|n \rangle$ is $\frac{1}{\sqrt{2
\pi}} e^{in\theta}$. In another physical explanation, $|n \rangle$
is seen as a site on an one dimensional lattice. $V$ is treated as
the kinetic energy which causes the electron hopping between
different sites and $H_0$ the potential energy of the electron in
the linear electric field. The Hamiltonian of the quantum kicked
rotator
\begin{equation}
H=-\frac{1}{2}\hbar^{2}\frac{\partial^2}{\partial\theta^{2}}-k
\cos\theta\sum_{n=1}^\infty \delta(t-n\tau)
\end{equation}
is just a lattice in a harmonic potential $V(x)=\frac{1}{2}\hbar^{2}
x^2$. The kick strength $k$ is the free diffusion time of the
electron.

In the field of quantum chaos, the term ``dynamic localization''
\cite{Casati1979, Fishman1982} means the absence of diffusion in the
momentum space of the quantum kicked rotator when the kick frequency
and the rotator frequency is incommensurable. Dunlap \textit{et al}
also used the term ``dynamic localization'' in the Bloch
oscillations problem (See Section V) \cite{Dunlap1986}. The result
of the localization of the original Maryland model (See Section IV)
is unexpected from the perspective of quantum chaos. But it is just
Bloch oscillations. We use the term Bloch oscillations in a broader
sense, while the ordinary meaning is an lattice electron in a time
independent linear electric field. Even if the hopping matrix
elements between different sites are time dependent, under some
condition, the electron still oscillates on the lattice. The
localization mechanisms of quantum kicked rotator \cite{Fishman1982}
and the original Maryland model and single band Bloch oscillations
are fundamentally different.

\section{\label{sec:level1}Exact solutions in the rotator representation}
In Section I, we solve the Maryland model in the site
representation, now we solve it in the rotator representation. The
Hamiltonian of Eq. $(1)$ in the rotator representation is
\begin{equation}
H=-i\omega(t) \hbar \frac{\partial}{\partial\theta}+V(\theta,t),
\end{equation}
where the unperturbed Hamiltonian
\begin{equation}
H_0=-i\omega(t) \hbar \frac{\partial}{\partial\theta} = \omega(t)p,
\end{equation}
where the angular momentum operator $p=-i \hbar
\frac{\partial}{\partial\theta}$. Since
\begin{equation}
\exp\bigg\{-\frac{i}{\hbar}\int_{0}^{t} \omega(t') \, dt' p\bigg\}
\end{equation}
is a translation operator, in the interaction picture,
\begin{equation}
\begin{split}
H_I(t)&=e^{\frac{i}{\hbar}\int_{0}^{t} \omega(t') \, dt'
p} V(\theta,t) e^{-\frac{i}{\hbar}\int_{0}^{t} \omega(t') \, dt' p} \\
&=V\bigg(\theta+\int_{0}^{t} \omega(t') \, dt', t\bigg) .
\end{split}
\end{equation}
$H_I(t)$ is a function of position $\theta$, so
$[H_I(t),H_I(t')]=0$. So the time ordering operator can also be
removed here and
\begin{equation}
U_I(t,0)=\exp\bigg\{-\frac{i}{\hbar} \int_{0}^{t}
V[\theta+\int_{0}^{t'} \omega(t'') \, dt'', t'] \, dt'\bigg\}.
\end{equation}
The unitary operator in Schrodinger picture is
\begin{equation}
\begin{split}
&U_S(t,0)=e^{\frac{i}{\hbar}\int_0^t H_0 \, dt} U_I(t,0) \\
&=e^{-\frac{i}{\hbar} \int_{0}^{t} \omega(t') \, dt'p} \times\\
&\qquad \exp\bigg\{-\frac{i}{\hbar} \int_{0}^{t}
V[\theta+\int_{0}^{t'} \omega(t'') \, dt'', t'] \, dt'\bigg\}.
\end{split}
\end{equation}
Since
\begin{equation}
\begin{split}
(H_I(t))_{nm} &=\frac{1}{2\pi}\int_{0}^{2\pi}
V\bigg(\theta+\int_{0}^{t} \omega(t') \, dt', t\bigg)
e^{i(m-n)\theta} \, d\theta\\
&=V_{nm} e^{i(n-m)(\int_{0}^{t} \omega(t') \, dt')},
\end{split}
\end{equation}
where $V_{nm}=V_{n-m}=\frac{1}{2\pi}\int_{0}^{2\pi} V(\theta, t)
e^{i(m-n)\theta} \, d\theta$. The equivalence between the methods in
Section I and this Section can be easily verified.

In a $M$ dimensional space, the solution of the Maryland model
\begin{equation}
H=\sum_{m=1}^{M} \omega_m(t) p_m+V(x_1,x_2,\cdots,x_M,t)
\end{equation}
is
\begin{equation}
\begin{split}
&U(t,0)=e^{-\frac{i}{\hbar} \int_{0}^{t} \sum_{m=1}^{M} \omega_m(t') p_m\, dt'} \times\\
&  \exp\bigg\{-\frac{i}{\hbar} \int_{0}^{t} V[x_1+\int_{0}^{t'}
\omega_1(t'') \, dt'', x_2+\int_{0}^{t'}
\omega_2(t'') \, dt'', \\
&\qquad \qquad \qquad \cdots, x_M+\int_{0}^{t'} \omega_M(t'') \,
dt'', t'] \, dt' \bigg\} ,
\end{split}
\end{equation}
where $x_m$ is the position and the momentum $p_m=-i \hbar
\frac{\partial}{\partial x_m}$.

A special case of Eq. $(17)$ and $(18)$ is Luttinger model
\cite{Luttinger1963, Mattis1965}
\begin{equation}
H=\sum_{n=1}^{N} p_{1,n}+\sum_{m=1}^{M}-p_{2,m}+\sum_{n=1}^{N}
\sum_{m=1}^{M} V(x_{1,n}-x_{2,m}).
\end{equation}
\begin{equation}
\begin{split}
U(t,0)&=\exp \bigg\{-\frac{i}{\hbar} (\sum_{n=1}^{N}p_{1,n}+\sum_{m=1}^{M}-p_{2,m}) t \bigg\} \times \\
&\qquad \exp \bigg\{-\frac{i}{\hbar} (\sum_{n=1}^{N} \sum_{m=1}^{M}
V(x_{1,n}-x_{2,m}))t \bigg\} ,
\end{split}
\end{equation}
where $x_{1,n}$, $p_{1,n}=-i\hbar\frac{\partial}{\partial x_{1,n}}$
are position and momentum of the $n$-th of ``1'' particles
(electrons) and $x_{2,n}$, $p_{2,m}=-i\hbar\frac{\partial}{\partial
x_{2,m}}$ are position and momentum of the $m$-th of ``2'' particles
(holes) and $N$ and $M$ are the total number of ``1'' and ``2''
particles respectively. The linear Toeplitz structure of Hamiltonian
in Eq. $(19)$ may be the origin of anomalous properties of Luttinger
model and Luttinger liquid compared with Fermi liquid
\cite{Luttinger1963, Mattis1965, Haldane1981}.

\section{\label{sec:level1}the original Maryland model}
The Hamiltonian of the original Maryland model or quantum kicked
linear rotator (QKLR) is
\begin{equation}
H=-i\hbar \frac{\partial}{\partial \theta}+ V(\theta)
\sum_{n=1}^\infty \delta(t-n\tau),
\end{equation}
where $\theta$ is the angle of the rotator, $\tau$ the kick period
and $k$ the kick strength. $p=-i \hbar \frac{\partial}{\partial
\theta}$ is the angular momentum and $V(\theta)=k \cos \theta$
\cite{Grempel1982, Prange1982, Berry1984}. QKLR is unphysical
because there is not a ground state. The classical version of Eq.
$(21)$ is not chaotic. The phase space is filled with invariant
curves \cite{Berry1984}. Does QKLR delocalize if $\frac{\tau}{2\pi}$
is rational? Berry proposed the energy of the rotator grows
quadratically \cite{Berry1984}. From the perspective of the exactly
solved eigenstates of the Floquet operator \cite{Grempel1982,
Prange1982, Fishman1982}, the extended eigenstates generally mean
delocalization. But we will prove in the section, QKLR always
localizes except one case when $\tau$ is an integer multiple of
$2\pi$. We set $\hbar=1$ and restrict $\tau$ to the domain
$(0,2\pi)$. The Floquet operator
\begin{equation}
F=e^{-iV(\theta)} e^{-i p \tau}.
\end{equation}
The matrix representation of the QKLR Hamiltonian is
\begin{equation}
H=\left(
\begin{array}{lllllll}
 \cdots & \cdots  &    &   &   &   &   \\
 \cdots & -2     &  \frac{k}{2}f&   &   &   &   \\
          &  \frac{k}{2}f    & -1 & \frac{k}{2}f &   &   &   \\
            &        &  \frac{k}{2}f & 0 & \frac{k}{2}f &   &   \\
          &        &    & \frac{k}{2}f& 1 & \frac{k}{2}f &   \\
            &        &    &   & \frac{k}{2}f & 2 & \cdots \\
            &        &    &   &   & \cdots & \cdots
\end{array}
\right),
\end{equation}
where $f=f(t)=\sum_{n=1}^\infty \delta(t-n\tau)$. In the interaction
picture,
\begin{equation}
\begin{split}
&H_I(t)=\left(
\begin{array}{llllll}
 \cdots & \cdots &   &   &   &   \\
 \cdots & 0 & \frac{k}{2} e^{-i t} f &   &   &   \\
   & \frac{k}{2} e^{i t} & 0 & \frac{k}{2} e^{-i t} f &   &   \\
   &   & \frac{k}{2} e^{i t} f & 0 & \frac{k}{2} e^{-i t} f &   \\
   &   &   & \frac{k}{2} e^{i t} f & 0 & \cdots \\
   &   &   &   & \cdots & \cdots
\end{array}
\right).
\end{split}
\end{equation}
$H_I(t)$ is a Laurent matrix. $[H_I(t_1), H_I(t_2)]=0$. From Eq.
$(8)$, to calculate $ U_I(t,0)=\exp[-i \int_0^t H_I(t') \, dt']$, we
need to calculate $\int_0^t H_I(t') \, dt'$. When $t=N\tau$, the
only non-zero matrix elements of $\int_0^{N\tau} H(t') \, dt'$ are
\addtocounter{equation}{1}
\begin{equation}
\begin{split}
&\bigg(\int_0^{N\tau} H(t') \, dt'\bigg)_{n+1,n}\\&=\frac{k}{2}
\frac{e^{i\tau}(1-e^{iN\tau})}{1-e^{i\tau}}=
\frac{k}{2}\frac{\sin(\frac{N\tau}{2})}{\sin(\frac{\tau}{2})}e^{i(N+1)\frac{\tau}{2}}=\gamma
e^{i\delta},
\end{split}
\end{equation}
where $\gamma=\frac{k}{2}\frac{\sin(N\tau/2)}{\sin(\tau/2)}$ and
$\delta=(N+1)\tau/2$, and
\begin{equation}
\bigg(\int_0^{N\tau} H(t') \, dt'\bigg)_{n,n+1}=\gamma e^{-i\delta}.
\end{equation}
Since $\gamma \leq \frac{k}{2 \sin(\tau/2)}$, $U(t,0)$ is a (almost)
band matrix. The rotator localizes, whether $\frac{\tau}{2\pi}$ is
rational or irrational except when $\tau$ is an integer multiple of
$2\pi$. But the localization mechanism is different from quantum
kicked rotator \cite{Fishman1982} as we discussed in Section II.

From the Appendix B,
\begin{equation}
U(N\tau,0)_{Inm}=e^{-i(m-n)\delta} i^{m-n}J_{n-m}(2\gamma).
\end{equation}
In Schrodinger picture,
\begin{equation}
U(N\tau,0)_{Snm}=e^{-inN\tau} e^{-i(m-n)\delta}
i^{m-n}J_{n-m}(2\gamma).
\end{equation}
If $\frac{\tau}{2 \pi}= \frac{p}{q}$, the period of QKLR is $q$
kicks ($2\pi p$). In quantum kicked rotator, $F=\exp(-i k
\cos\theta) \exp(-i p^2 \frac{\tau}{2})$. When $\tau=2\pi$, $F$ is
the same with QKLR with $\tau=\pi$. So the period of quantum kicked
rotator with $\tau=2\pi$ is $2\tau$ ($4\pi$) \cite{Casati1979}.

Now we discuss the seemingly conflict between the unitary operator
$F^N$ and eigenstates of the Floquet operator $F$ calculated in
\cite{Grempel1982, Prange1982, Berry1984}. We calculate eigenvalues
and eigenstates using a method similar to \cite{Berry1984}. The
eigenvalue equation of the Floquet operator $F$ is
\begin{equation}
F \phi(\theta) =e^{-iV(\theta)} e^{-i p \tau} \phi(\theta)= \lambda
\phi(\theta).
\end{equation}
Since $e^{-i p \tau}$ is a translation operator,
\begin{equation}
e^{-iV(\theta)} e^{-i p \tau} \phi(\theta)= e^{-iV(\theta)}
\phi(\theta-\tau)=\lambda \phi(\theta).
\end{equation}
We assume $\frac{\tau}{2\pi}=\frac{p}{q}$. From Eq. $(31)$,
\begin{eqnarray}
e^{-iV(\theta-\tau)} \phi(\theta-2\tau)&=&\lambda \phi(\theta-\tau); \nonumber\\
e^{-iV(\theta-2\tau)} \phi(\theta-3\tau)&=&\lambda \phi(\theta-2\tau); \nonumber\\
&\vdots & \nonumber\\
e^{-iV(\theta-(q-1)\tau)} \phi(\theta-q\tau)&=&\lambda
\phi(\theta-(q-1)\tau).
\end{eqnarray}
From Eq. $(31)$, $(32)$ and $\phi(\theta-q\tau)=\phi(\theta)$,
\begin{equation}
\exp(-i \sum_{n=0}^{q-1} V(\theta-n \tau))=\lambda^q.
\end{equation}
\begin{equation}
\begin{split}
\sum_{n=0}^{q-1} V(\theta-n \tau)&=\sum_{n=0}^{q-1} k\cos(\theta-n
\tau)\\
&=\cos(\theta+(q-1)\frac{\tau}{2})\frac{\sin(\frac{q\tau}{2})}{\sin(\frac{\tau}{2})}\\
&=0.
\end{split}
\end{equation}
Another derivation of Eq. $(34)$ is
\begin{equation}
\begin{split}
\sum_{n=0}^{q-1} k\cos(\theta-n \tau)&=k Re\bigg[\sum_{n=0}^{q-1}
\exp(i(\theta-n\tau))\bigg]\\
&=k Re\bigg[\exp(i\theta) \sum_{n=0}^{q-1} \exp(i(-n\tau))\bigg]\\
&=0,
\end{split}
\end{equation}
where $Re$ is the real part of a complex number. So
\begin{equation}
\lambda^q=1.
\end{equation}
\begin{equation}
\lambda=1,e^{i \frac{2\pi}{q}}, e^{i 2 \frac{2\pi}{q}}, \cdots,e^{i
(q-1) \frac{2\pi}{q}}.
\end{equation}
The Floquet operator $F$ has only $q$ eigenvalues. $F$ for rational
$\frac{\tau}{2\pi}$ has infinite degenerate point spectra. Grempel
\textit{et al} thought quasienergy bands have a finite width
\cite{Grempel1982}. Now we find eigenstates. Let's restrict $\theta$
to the domain $[0,\frac{2\pi}{q}]$. Define $\phi(\theta)$ in the
domain $[0,\frac{2\pi}{q}]$ as an arbitrary function and in other
domains
\begin{equation}
\begin{split}
\phi(\theta-\tau)&=\lambda e^{iV(\theta)} \phi(\theta); \\
\phi(\theta-2\tau)&=\lambda e^{iV(\theta-\tau)} \phi(\theta-\tau); \\
&\ \, \vdots \\
\phi(\theta-(q-1)\tau)&=\lambda
e^{iV(\theta-(q-2)\tau)}\phi(\theta-(q-2)\tau).
\end{split}
\end{equation}
Eq. $(38)$ gives the eigenstates of $F$.

When $\frac{\tau}{2\pi}$ is irrational, the eigenvalues are
$e^{il\tau}$, where $l$ is an arbitrary integer, and the eigenstates
are localized \cite{Grempel1982, Prange1982, Berry1984}. From Eq.
$(38)$, we can construct localized and extended eigenstates. There
are no real conflict between the unitary operator in Eq. $(28)$ and
$(29)$ and the extended eigenstates calculated in \cite{Grempel1982,
Prange1982, Berry1984}.

\section{\label{sec:level1}Time dependent $H_0$}
The time dependent $H_0$ considered by Dunlap \textit{et al}
\cite{Dunlap1986} is
\begin{equation}
\begin{split}
H(t)&=T \sum_{m=-\infty}^{\infty}(|m\rangle \langle m+1| + |m+1
\rangle \langle m|)\\
&\quad+E(t) \sum_{m=-\infty}^{\infty} m|m \rangle \langle m|,
\end{split}
\end{equation}
where $T$ is the nearest-neighbor coupling and $E(t)$ is the time
dependent linear electric potential. Dunlap \textit{et al} gave the
analytic solution of the above Hamiltonian \cite{Dunlap1986} and
found the time dependent field $H_0$ generally destroys Bloch
oscillations.. Here we treat Eq. $(39)$ as a Maryland model or a
linear Toeplitz system.
\begin{equation}
\begin{split}
H(t)&=\sum_n \sum_{m=-\infty}^{\infty} T_{n}(t)|m+n \rangle \langle
m|\\
&\qquad+E(t) \sum_{m=-\infty}^{\infty} m|m \rangle \langle m|.
\end{split}
\end{equation}
In the interaction picture,
\begin{equation}
\begin{split}
H_I(t) =\sum_n \sum_{m=-\infty}^{\infty} T_{n}(t) e^{i n \int_0^{t}
E(t')\, dt'} |m+n \rangle \langle m|.
\end{split}
\end{equation}
$H_{I}(t)$ is a Laurent matrix. $U(t,0)_I=\exp[-i \int_0^t H_I(t')
\, dt']$. In the simple case of \cite{Dunlap1986},
$T_{1}(t)=T_{-1}(t)=T$ and $E(t)=-E \sin t$. In $N$ periods from $0$
to $2N\pi$, for $n=1$,
\begin{equation}
\begin{split}
&\int_{0}^{2N\pi} T_1(t') e^{i \int_0^{t'} E(t'') \, dt''} \, dt'\\
&=T \int_{0}^{2N\pi} e^{i \int_0^{t'} -E \sin t'' \, dt''} \, dt'=2
\pi N T J_0(E)e^{-iE};
\end{split}
\end{equation}
and for $n=-1$,
\begin{equation}
\int_{0}^{2N\pi} T_{-1}(t') e^{-i \int_0^{t'} E(t'') \, dt''} \,
dt'=2 \pi N T J_0(E)e^{iE}.
\end{equation}
In the derivation of Eq. $(42)$ and $(43)$, we used the formula
\begin{equation}
e^{iz\cos\theta}=\sum_{n=-\infty}^{\infty}J_n(z)i^n e^{in\theta},
\end{equation}
of \cite{Abramowitz1972, ZhuxiWang}. From the Appendix B, the matrix
elements of the unitary operator in the interaction picture is
\begin{equation}
U(t,0)_{Inm}=e^{i(m-n)E}
i^{m-n}J_{n-m}(4 \pi N T J_0(E)).
\end{equation}
Since
\begin{equation}
e^{i \int_0^{2 \pi N} H_0(t') \, d t'}=1,
\end{equation}
the unitary operator in Schrodinger picture
\begin{equation}
U(2N\pi,0)_S=U(2N\pi,0)_I.
\end{equation}

From Eq. $(46)$ and $(47)$, the electron will diffuse away except
when $J_0(E)=0$. In the case of $J_0(E)=0$, the electron will not
delocalize. This is referred as dynamic localization by Dunlap
\textit{et al} \cite{Dunlap1986}.

\section{\label{sec:level1}Conclusion and Discussion}
In summary, the Maryland model with doubly infinite Hilbert space is
exactly solved. In the interaction representation, the unitary
operator is the matrix exponential of a Laurent matrix. It is the
special structure renders the Hamiltonian solvable. We think the
solution can be generalized to a more general structure of
Hamiltonian. We give the correct solution of the original Maryland
model, concerning the resonant cases. Compared with Dunlap
\textit{et al}'s method \cite{Dunlap1986} to solve the lattice
electron in a time dependent linear electric field, our method based
on the Maryland model is physically appealing and simpler. Further
work should generalize the structure of linear Toeplitz system and
remove the requirement of doubly infinite Hilbert space.

\appendix
\section{A theorem of Laurent matrix}
``Doubly infinite dimensional Toeplitz matrix'' is referred as
Laurent matrix in the mathematical literature. Two Laurent matrices
commute. In the appendix, we give a proof (it may exist in another
place) of the commutativity. A Laurent matrix $A$ is defined as
$A_{nm}=A_{n+i,m+i}=A_{n-m}$, where $i$ is an arbitrary integer and
$n,m$ run from $-\infty$ to $\infty$. $C=AB$.
$C_{nm}=\sum_l{A_{nl}B_{lm}}$. $D=BA$.
$D_{nm}=\sum_{p}{B_{np}A_{pm}}$.

\begin{equation}
\begin{split}
C_{nm}-D_{nm}&=\sum_l{A_{nl}B_{lm}}-\sum_p{B_{np}A_{pm}} \\
&=\sum_l{A_{n-l}B_{l-m}}-\sum_p{B_{n-p}A_{p-m}} \\
&=\sum_l{A_{n-l}B_{l-m}}-\sum_p{A_{p-m}B_{n-p}}\\
\end{split}
\end{equation}
Given a $l$, there is a $p=n+m-l$, which satisfies $n-l=p-m$ and
$l-m=n-p$, and \textit{vice versa}. So $\sum_l{A_{n-l}B_{l-m}}$ and
$\sum_p{A_{p-m}B_{n-p}}$ contain the same terms. $C_{nm}-D_{nm}=0$.
$C=D$. $[A,B]=0$. Note the condition of doubly infinite dimension is
necessary to ensure $p=n+m-l$ is always the index of an existent
matrix element.

In fact, the above theorem is trivial because every Laurent matrix
$A$ is a function of the position $\theta$ (a multiplication
operator) in the position representation.
\begin{equation}
A(\theta)= \sum_{n=-\infty}^{\infty}A_n e^{-i n \theta}.
\end{equation}
Two functions of position commute with each other.

\section{Matrix exponential of a Laurent matrix}
We now calculate the matrix exponential $e^{-iM}$ of a simple
Laurent Hermitian matrix $M$. $M$ is bidiagonal; only $M_{n+1,n}$
and $M_{n,n+1}$ are not zero. \addtocounter{equation}{1}
\begin{equation}
\begin{split}
&M_{n+1,n}=\gamma e^{i\delta}; \\
&M_{n,n+1}=\gamma e^{-i\delta},
\end{split}
\end{equation}
where $\gamma$ and $\delta$ are real numbers.

If we choose the basis of $M$ as $\frac{1}{\sqrt{2\pi}}e^{i n
\theta}$, in the rotator representation,
\begin{equation}
\begin{split}
M(\theta)=\gamma e^{i\delta} e^{i\theta}+\gamma e^{-i\delta}
e^{-i\theta} =2\gamma \cos(\theta+\delta).
\end{split}
\end{equation}

The matrix element of $e^{-iM}$ is
\begin{equation}
\begin{split}
&(e^{-iM})_{nm}\\
&=\frac{1}{2\pi}\int_0^{2\pi}e^{-iM(\theta)}e^{i(m-n)\theta} \, d\theta \\
&=\frac{1}{2\pi} \int_0^{2\pi}e^{-i2 \gamma
\cos(\theta+\delta)}e^{i(m-n)\theta} \, d\theta\\
&=\frac{1}{2\pi} \int_0^{2\pi}e^{-i2 \gamma
\cos(\theta+\delta)}e^{i(m-n)(\theta+\delta)}e^{-i(m-n)\delta} \, d\theta\\
&=e^{-i(m-n)\delta} i^{n-m}J_{n-m}(-2\gamma)\\
&=e^{-i(m-n)\delta} i^{n-m}(-1)^{n-m}J_{n-m}(2\gamma)\\
&=e^{-i(m-n)\delta} i^{m-n}J_{n-m}(2\gamma),
\end{split}
\end{equation}
where $J$ is the Bessel function of the first kind. In the
derivation of Eq. $(B4)$, we used Eq. $(44)$ and
\begin{equation}
J_n(z)=(-1)^n J_n(-z)
\end{equation}
of \cite{Abramowitz1972, ZhuxiWang}.
\bibliography{Linear}
\end{document}